\documentclass[aps,twocolumn,groupedaddress,amsmath,amssymb]{revtex4}
\usepackage{graphicx}
\usepackage{dcolumn}
\usepackage{bm}
\usepackage{amssymb}
\usepackage{amsmath}
\usepackage{epsfig}    
\usepackage{color}
\usepackage{slashed}
\usepackage{hhline}

\def\be{\begin{equation}}
\def\ee{\end{equation}}
\newcommand{\bea}{\begin{eqnarray}}
\newcommand{\eea}{\end{eqnarray}}
\newcommand{\nn}{\nonumber}

\begin{document}


\title{An inverse seesaw model with
$U(1)_R$ gauge symmetry
}

\author{Takaaki Nomura}
\email{nomura@kias.re.kr}
\affiliation{School of Physics, KIAS, Seoul 02455, Korea}

\author{Hiroshi Okada}
\email{hiroshi.okada@apctp.org}
\affiliation{Asia Pacific Center for Theoretical Physics, Pohang, Geyoengbuk 790-784, Republic of Korea}

\date{\today}

\begin{abstract}
 We propose a natural realization of inverse seesaw model with right-handed and flavor dependent $U(1)$ gauge symmetries, in which we formulate the neutrino mass matrix to reproduce current neutrino oscillation data in a general way. Also we study a possibility to provide predictions to the neutrino sector by imposing an additional flavor dependent $U(1)_{L_\mu-L_\tau}$ gauge symmetry that also satisfies the gauge anomaly cancellation conditions associated with $U(1)_R$.  Then we analyze collider physics on an extra gauge boson and show a possibility of detection. 
 \end{abstract}
\maketitle

\section{Introduction}
$U(1)_{B-L}$~\cite{Mohapatra:1980qe} and $U(1)_R$~\cite{Jung:2009jz, Ko:2013zsa, Nomura:2016emz,Nomura:2017tih,Chao:2017rwv} symmetries require three families of neutral right-handed (or left-handed) fermions in order to cancel the gauged anomalies.
A feature of these two symmetries  is very similar each other about Yukawa sector.
In fact, once a lepton Yukawa model is constructed in a gauge symmetry, the other symmetry can also reproduce the same one.
And these symmetries are known as a natural extension of the standard model (SM) to realize various seesaw mechanisms
such as canonical seesaw model~\cite{Seesaw1, Seesaw2, Seesaw3, Seesaw4}, inverse seesaw model~\cite{Mohapatra:1986bd, Wyler:1982dd}, linear seesaw model~\cite{Wyler:1982dd, Akhmedov:1995ip, Akhmedov:1995vm}, etc. 

On the other hand, nature of these two gauge sectors are so different each other, and one might be able to test their differences via current or future experiments so as to make use of the polarized electron/positron beam at e.g., ILC~\cite{Barklow:2015tja}.  
Indeed, $U(1)_{B-L}$ is chirality-universal in a kinetic term, while $U(1)_R$ has right-handed chirality only.
 In this sense, it would be worthwhile for us  to construct models with gauged $U(1)_{B-L}$ and/or $U(1)_{R}$ symmetry as many as possible,
so that we can distinguish these two extra symmetries in a various phenomenological points of view. 

 In this paper, we construct an inverse seesaw model with $U(1)_R$ symmetry, in which we formulate the neutrino mass matrix to reproduce current neutrino oscillation data~\cite{pdg} in a general way.
 Inverse seesaw requires a left-handed neutral fermions $S_L$ in addition to the right-handed ones $N_R$, and provides us more
 complicated neutrino mass matrix. Therefore, each of mass hierarchies are softer than the other models such as canonical seesaw and it could provide abundant phenomenologies such as unitarity constraints.
 {\it
 Note that we expect $S_L$ has nonzero $U(1)_{B-L}$ charge, because it is a kind of partner of $N_R$.
 In that case, however, $U(1)_{B-L}$ can not be gauged since anomaly cancellation condition can not be satisfied.
 Therefore introduction of left-handed singlet fermion is more natural in gauged $U(1)_R$ symmetry case compared with gauged $U(1)_{B-L}$ symmetry case since the left-handed singlet fermion cannot have lepton number in the latter case.} 
Also we study a possibility to provide predictions to the neutrino sector by imposing an additional flavor dependent $U(1)_{L_\mu-L_\tau}$ gauge symmetry that also satisfies the gauge anomaly cancellations among $U(1)_R$.
\footnote{$U(1)_{B-L}\times U(1)_{L_\mu- L_\tau}$ can also be anomaly free. See ref.~\cite{Araki:2012ip}.}
 Then we analyze collider physics on an extra gauge boson and show a possibility of detection. 
 
This letter is organized as follows.
In Sec. II, {we review our model and formulate the lepton sector. 
Then we discuss phenomenologies of neutrinos and an extra neutral gauge boson at colliders.
 Finally we devote the summary of our results and the conclusion.}

\section{Model setup and Constraints}
\begin{table}[t!]
\begin{tabular}{|c||c|c|c|c|c|c|c||c|c|c|}\hline\hline  
& ~$Q_L^a$~& ~$u_R^a$~  & ~$d_R^a$~& ~$L_L^a$~& ~$e_R^a$~& ~$N_R^a$~& ~$S_L^a$~& ~$H$~& ~$\varphi_1$~& ~$\varphi_2$~\\\hline\hline 
$SU(3)_C$ & $\bm{3}$  & $\bm{3}$ & $\bm{3}$ & $\bm{1}$ & $\bm{1}$ & $\bm{1}$ & $\bm{1}$ & $\bm{1}$ & $\bm{1}$ & $\bm{1}$  \\\hline 
$SU(2)_L$ & $\bm{2}$  & $\bm{1}$  & $\bm{1}$  & $\bm{2}$  & $\bm{1}$  & $\bm{1}$  & $\bm{1}$ & $\bm{2}$   & $\bm{1}$ & $\bm{1}$   \\\hline 
$U(1)_Y$   & $\frac16$ & $\frac23$ & $-\frac13$ & $-\frac12$  & $-1$ & $0$  & $0$  & $\frac12$ & $0$  & $0$\\\hline
$U(1)_{R}$   & $0$ & $1$ & $-1$   & $0$  & $-1$  & $1$  & $0$  & $1$ & $1$  & $0$\\\hline
\end{tabular}
\caption{ 
Charge assignments of the our fields
under $SU(3)_C\times SU(2)_L\times U(1)_Y\times U(1)_{R}$, where the upper index $a$ is the number of family that runs over 1-3.  Singlet scalar $\varphi_2$ is required when we add $U(1)_{L_\mu -L_\tau}$ gauge symmetry.}
\label{tab:1}
\end{table}

In this section we formulate our model.
At first, we add three families of right(left)-handed fermions $N_R(S_L)$ with 1(0) charge under the $U(1)_R$ gauge symmetry,
and an isospin singlet boson $\varphi_1$ with 1 charge under the same symmetry.
Here we denote each of vacuum expectation value to be $\langle H\rangle\equiv v_H/\sqrt2$, and $\langle \varphi_1\rangle\equiv v_{\varphi_1}/\sqrt2$.
Furthermore, the SM Higgs boson $H$ also has 1 charge to induce the masses of SM fermions from the Yukawa Lagrangian after the spontaneously symmetry breaking.~\footnote{Due to the feature of nonzero charges of $H$, lower bound on the breaking scale of $U(1)_R$ is determined via the precision test of $Z$ boson mass; $\Lambda \gtrsim {\cal O}$(10) TeV~\cite{Nomura:2017tih}.} 
All the field contents and their assignments are summarized in Table~\ref{tab:1}.
The relevant Yukawa Lagrangian under these symmetries is given by~\footnote{Since the quark sector is exactly same as the one of SM, we neglect it hereafter.} 
\begin{align}
&-{\cal L_\ell}
=  y_{\ell_{aa}} \bar L^a_L H e^a_R  +  y_{D_{ab}} \bar L^a_L\tilde H N^b_R  +  y_{{SN}_{aa}} \bar S^{a}_L N^a_R \varphi^*_1\nn\\
&+\mu_{ab} \bar S^a_L (S^c_L)^{b}
+ {\rm h.c.}, \label{Eq:yuk}
\end{align}
where $\tilde H\equiv i\sigma_2H$, and upper indices $(a,b)=1$-$3$ are the number of families, and $y_\ell$ and $y_{SN}$ can be diagonal matrix without loss of generality due to the redefinitions of the fermions.
Each of the mass matrix is defined by $m_\ell=y_\ell v_H/\sqrt2$, $m_D=y_D v_H/\sqrt2$, and  $M_{SN}=y_{SN} v_{\varphi_1}/\sqrt2$.   
 Notice that $S_L$ is singlet under all gauge symmetry and it does not interact with any gauge interactions without mixing among neutral fermions. Here we denote $S_L$ as left-handed in a sense it dose not have $U(1)_R$ charge. As we discuss below heavy extra neutral fermion mass is approximately given by $M_{NS}$ which is taken to be TeV scale. In addition, Majorana mass term of $S_L$ breaks lepton number as we assign lepton number to $S_L$.

\subsection{Neutrino sector without $U(1)_{L_\mu-L_\tau}$}
After the spontaneously symmetry breaking, neutral fermion mass matrix with 9$\times$9 is given by
\begin{align}
M_N
&=
\left[\begin{array}{ccc}
0 & m_D & 0  \\ 
m_D^T & 0 & M_{NS}^T \\ 
0  & M_{NS} & \mu \\ 
\end{array}\right].
\end{align}
Then the active neutrino mass matrix can approximately be found as
\begin{align}
m_\nu\approx m_D M_{NS}^{-1} \mu (M_{NS}^T)^{-1} m_D^T,
\end{align}
where $\mu<< m_D\lesssim M_{NS}$ is expected
\footnote{These hierarchies could be explained by several mechanisms such as radiative models~\cite{Dev:2012sg, Dev:2012bd, Das:2017ski} and effective models with higher order terms \cite{Okada:2012np}.}.
The neutrino mass matrix is diagonalized by unitary matrix $U_{MNS}$; $D_\nu= U_{MNS}^T m_\nu U_{MNS}$, where $D_\nu\equiv {\rm diag}(m_1,m_2,m_3)$. 
One of the elegant ways to reproduce the current neutrino oscillation data~\cite{pdg} is to apply the Casas-Ibarra parametrization~\cite{Casas:2001sr}
without loss of generality, and find the following relation
\begin{align}
m_D=U_{MNS}^* \sqrt{D_\nu} O_{mix} \sqrt{I_N} (L^T_N)^{-1}.
\end{align}
Here $O_{mix}$ is an arbitrary 3 by 3 orthogonal matrix with complex values, $I_N$ is a diagonal matrix, and $L_N$ is a lower unit triangular~\cite{Baek:2017qos}, which can uniquely be decomposed to be $M_{NS}^{-1} \mu (M_{NS}^T)^{-1}=L_N^T I_N L_N$,
since it is symmetric matrix. Note here that all the components of $m_D$ should not exceed 246 GeV, once perturbative limit of $y_D$ is taken to be 1. 

Mass scale of heavy neutral fermions is approximately given by $M_{NS}$ which is taken to be $\mathcal{O}(1)$ TeV in our scenario. 
Then neutrino mass scale is 
\begin{equation}
m_\nu \sim  10^{-6} \left( \frac{m_D}{\rm GeV} \right)^2 \left( \frac{M_{NS}}{\rm TeV} \right)^2 \frac{\mu}{\rm GeV} \ {\rm GeV}.
\end{equation}
Thus we can realize neutrino mass scale $\sim 0.1$ eV with $\mu \sim 1 (0.0001)$ GeV for $m_D \sim 0.01 (1)$ GeV.
In addition, new fermions are not decoupled at TeV scale even if scale of $v'$ is as large as $\gtrsim 18 $ TeV.

\subsection{Neutrino sector with $U(1)_{L_\mu-L_\tau}$}
Here we introduce local $U(1)_{L_\mu-L_\tau}$ symmetry to restrict neutrino mass structure in inverse seesaw scenario~\cite{Biswas:2018yus, Dev:2017fdz} 
 where we add SM singlet scalar $\varphi_2$ with $L_\mu -L_\tau$ charge $1$ to break the symmetry spontaneously.  
Then Yukawa interactions and Majorana masses are constrained, and we have new Yukawa interactions; 
\begin{equation}
-\mathcal{L}_{\rm new} = y_{ij} \varphi_2 \bar S^i_L (S_L^c)^j + h.c.,
\end{equation}
where index $i(j)$ is determined to satisfy gauge invariance.
Thus once we impose $U(1)_{L_\mu-L_\tau}$ gauge symmetry as shown in table~\ref{tab:2}
\footnote{Before the discussion of neutrino sector, we have to check the gauge anomalies.
The non-trivial gauge anomalies are $[U(1)_{R}]^2 U(1)_{L_\mu-L_\tau}$ and $U(1)_{R} [U(1)_{L_\mu-L_\tau}]^2$, and one straightforwardly confirms
that there are no anomalies in our field assignments.},
the mass matrices $m_D,M_{SN},\mu$ are specified to be 
\begin{align}
M_{SN}
&=
\left[\begin{array}{ccc}
m_{SN_1} & 0 & 0  \\ 
0 & m_{SN_2} & 0 \\ 
0  & 0 & m_{SN_3} \\ 
\end{array}\right], \nn \\
m_D
&=
\left[\begin{array}{ccc}
m_{d_1} & 0 & 0  \\ 
0 & m_{d_2} & 0 \\ 
0  & 0 & m_{d_3} \\ 
\end{array}\right],\quad
\mu
=
\left[\begin{array}{ccc}
\mu_{1} & \mu_2 & \mu_3  \\ 
\mu_2 & 0 & \mu_4 \\ 
\mu_3 & \mu_4 & 0 \\ 
\end{array}\right], 
\end{align}
where $\mu_{2,3}$ is induced only after the $U(1)_{L_\mu-L_\tau}$ spontaneously symmetry breaking.
Therefore, the neutrino mass matrix directly reflects the form of $\mu$ as 
\begin{align}
m_\nu
&=
\left[\begin{array}{ccc}
\mu_1 \frac{m_{d_1}^2}{m_{SN_1}^2} & \mu_2 \frac{m_{d_1} m_{d_2}}{m_{SN_1}m_{SN_2}} & \mu_3 \frac{m_{d_1} m_{d_3}}{m_{SN_1}m_{SN_3}}   \\ 
\mu_2 \frac{m_{d_1} m_{d_2}}{m_{SN_1}m_{SN_2}}  & 0 & \mu_4 \frac{m_{d_2} m_{d_3}}{m_{SN_2}m_{SN_3}}  \\ 
 \mu_3 \frac{m_{d_1} m_{d_3}}{m_{SN_1}m_{SN_3}}  & \mu_4 \frac{m_{d_2} m_{d_3}}{m_{SN_2}m_{SN_3}} & 0 \\ 
\end{array}\right].
\end{align}
Thus we can predict inverted neutrino ordering and specific value of Dirac phase by analyzing the two-zero texture~\cite{Fritzsch:2011qv,Dev:2017fdz}. 
Here the number of parameters in the neutrino mass matrix is nine real parameters (that are equivalent of four complexes and one real). Then one more phase is there in addition to the Dirac phase and two Majorana phases.
{\it Note here that this two-zero texture originates from $\mu$ in the inverse seesaw model that cannot be reproduced by a canonical seesaw model.}

\begin{table}[t!]
\begin{tabular}{|c||c|c|}\hline\hline  
&~$L_L^2,N_R^2, S_L^2, e_R^2,\varphi_2$~&  ~$L_L^3,N_R^3, S_L^3, e_R^3$~~\\\hline\hline 
$U(1)_{L_\mu-L_\tau}$  & $1$ & $-1$\\\hline
\end{tabular}
\caption{ 
Charge assignments of our fields
under $ U(1)_{L_\mu-L_\tau}$, where the other fields do not have $L_\mu - L_\tau$ charge.}
\label{tab:2}
\end{table}

\subsection{Non-unitarity}
Here, let us briefly discuss non-unitarity matrix $U'_{MNS}$.
This is typically parametrized by the form 
\begin{align}
U'_{MNS}\equiv \left(1-\frac12 FF^\dag\right) U_{MNS},
\end{align}
where $F\equiv  (m^{T}_{NS})^{-1} m_D$ is a hermitian matrix, and $U'_{MNS}$ represents the deviation from the unitarity. 
The global constraints are found via several experimental results such as the SM $W$ boson mass $M_W$, the effective Weinberg angle $\theta_W$, several ratios of $Z$ boson fermionic decays, invisible decay of $Z$, electroweak universality, measured Cabbibo-Kobayashi-Maskawa, and lepton flavor violations~\cite{Fernandez-Martinez:2016lgt}.
The result is then given by~\cite{Agostinho:2017wfs}
\begin{align}
|FF^\dag|\le  
\left[\begin{array}{ccc} 
2.5\times 10^{-3} & 2.4\times 10^{-5}  & 2.7\times 10^{-3}  \\
2.4\times 10^{-5}  & 4.0\times 10^{-4}  & 1.2\times 10^{-3}  \\
2.7\times 10^{-3}  & 1.2\times 10^{-3}  & 5.6\times 10^{-3} \\
 \end{array}\right].
\end{align} 
Once we conservatively take $F\approx 10^{-5}$, we find $\mu\approx$1-10 GeV to satisfy the typical neutrino mass scale,
which can be easy to realize. 
 In addition, for $M_{NS} \sim 1$ TeV, we require $y_D \sim 10^{-4}$ which is slightly larger than case of Type-I seesaw~\cite{Abbas:2007ag}.

\subsection{Collider physics}

Here we discuss collider physics of our model mainly focusing on $Z'_R$ boson from $U(1)_R$ which obtain its mass via the vacuum expectation value of $\varphi_2$.
The gauge interaction associated with $Z'_R$ is given by
\begin{align}
\mathcal{L} \supset g_R \left( \bar u_R \gamma_\mu u_R - \bar d_R \gamma_\mu d_R - \bar \ell_R \gamma_\mu \ell_R + \bar N_R \gamma_\mu N_R  \right) Z'^\mu_R, 
\end{align}
where $g_R$ is gauge coupling constant for $U(1)_R$, and flavor index is omitted.

\begin{figure}[tbh]
\begin{center}
\includegraphics[width=70mm]{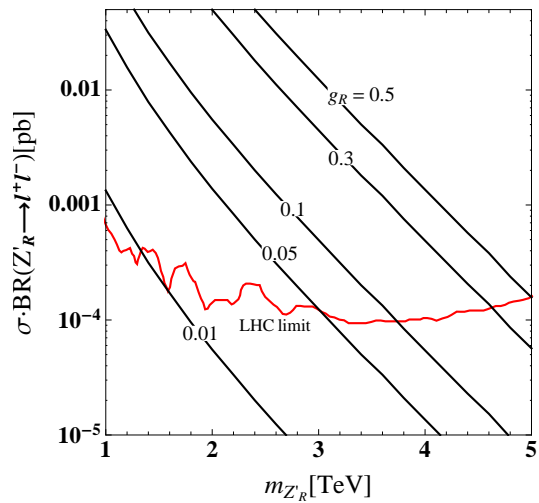} \qquad
\caption{The product of $Z'_R$ production cross section and $BR(Z'_R \to \ell^+ \ell^-)$ where region above red curve is excluded by the latest data~\cite{Aaboud:2017buh}. } 
  \label{fig:ZpLHC}
\end{center}\end{figure}

{\it $Z'_R$ physics at the LHC} : In our model $Z'_R$ can be produced via $q \bar q \to Z'_R$ process, and it will decay into SM fermions and $N_R$ if kinematically allowed.
Then stringent constraint is given by di-lepton resonance search at the LHC.
We estimate the cross section with {\it CalcHEP}~\cite{Belyaev:2012qa} by use of the CTEQ6 parton distribution functions (PDFs)~\cite{Nadolsky:2008zw}, implementing relevant interactions.
In addition, we find branching ratio (BR) for the decay mode $Z'_R \to e^+e^-/\mu^+ \mu^-$ is $\sim 4.8 \%$ for both electron and muon when we assume $\bar N_R N_R$ mode is not kinematically allowed;
even if we include $\bar N_R N_R$ mode the BR does not change much as $BR(Z'_R \to e^+ e^-/\mu^+\mu^-) \gtrsim 4.2 \%$. 
In Fig.~\ref{fig:ZpLHC}, we show $\sigma(pp \to Z'_R) BR(Z'_R \to \ell^+ \ell^-)$ as a function of $m_{Z'_R}$ for several values of $g_R$ where the $BR$ is sum of electron and muon mode and the red curve indicate the LHC limit obtained from ref.~\cite{Aaboud:2017buh}.
We find that $Z'_R$ mass should be heavier than $\sim 3.8$ TeV for $g_R = 0.1$ where corresponding production cross section is $\sigma(pp \to Z'_R) \lesssim 1$ fb. 

Here we discuss production of heavy neutrino $\nu_H$ at the LHC via $Z'_R$ boson.
If masses of $\nu_{H_i}$ are sufficiently lighter than $m_{Z'_R}/2$, $BR(Z'_R \to \nu_H \nu_H)$ is around $4 \%$ for each mass eigenstate.
Then $\nu_H$ decays such that $\nu_H \to W^\pm \ell^\mp$ and $\nu_H \to Z \nu_L$ via mixing in neutrino sector. 
As we discussed above, $Z'_R$ production cross section is less than $\sim 1$ fb for $m_{Z'_R}$ being several TeV scale,
and $\nu_H$ production cross section will be $\sigma \cdot BR \lesssim 0.04$ fb for each mass eigenstate.
Thus large integrated luminosity is required to obtain sufficient number of events to analyze the signal.
It is also important to confirm the ratio of the $BR$ of each decay mode of $Z'_R$ to distinguish it from other $Z'$ boson like that from $U(1)_{B-L}$ gauge symmetry where the approximated $BR$s are given in Table.~\ref{tab:BRs}. 

Here we also comment on $Z'_{\mu - \tau}$ boson from $U(1)_{L_\mu - L_\tau}$ gauge symmetry. 
Gauge interactions among $Z'_{\mu - \tau}$ and fermions are written by
\begin{align}
\mathcal{L} \supset & g_{\mu - \tau}  ( \bar \mu \gamma_\alpha \mu - \bar \tau \gamma_\alpha \tau 
+ \bar \nu_\mu \gamma_\alpha P_L \nu_\mu - \bar \nu_\tau \gamma_\alpha P_L \nu_\tau \nonumber \\
& \qquad + \bar N^2 \gamma_\alpha P_R N^2 - \bar N^3 \gamma_\alpha P_R N^3  \nonumber \\
& \qquad + \bar S^2 \gamma_\alpha P_L S^2 - \bar S^3 \gamma_\alpha P_L S^3 ) Z'^\alpha_{\mu -\tau},
\end{align}
where $g_{\mu-\tau}$ is the gauge coupling constant of $U(1)_{L_\mu - L_\tau}$ and fermions are flavor eigenstates.
It is difficult to detect $Z'_{\mu - \tau}$ when we consider it to be light as $\mathcal{O}(10)$-$\mathcal{O}(100)$ MeV so that muon $g-2$ can be explained~\cite{Gninenko:2001hx}.
The $Z'_{\mu - \tau}$ interaction induces flavor violating decay of heavy neutrino such as $\nu_{H_i} \to Z'_{\mu - \tau} \nu_{H_j}$ where $m_{\nu_{H_i}} > m_{\nu_{H_j}} $.
Thus phenomenology of heavy neutrino at the LHC would be affected by the gauge boson. However detailed analysis is beyond the scope of the paper. 
We note that the $Z'_{\mu - \tau}$ gauge interaction with charged lepton is flavor diagonal and do not induce any flavor violations(LFVs) even at loop levels, once we do not seriously take mixings among neutral gauge bosons into consideration. 
Contribution via $W^{\pm}$ and neutral fermions at one-loop level also gives no LFVs, considering negligible mixing among neutral fermions. 
Even when we consider the mixing among gauge bosons or neutral fermions, their mixings are so small because they are respectively proportional to $(m_{Z'_{\mu-\tau}}/m_{Z(Z'_R)})^2\lesssim 10^{-6}(10^{-8})$ and $(m_D/M_{SN})^2\lesssim 10^{-6}$~\cite{Khalil:2010iu}.

\begin{table}[t!]
\begin{tabular}{|c||c|c|c|}\hline\hline  
Mode & ~$\ell^- \ell^+$~ & ~$q \bar q$~ & ~$\nu_H \bar \nu_H$~ \\ \hline\hline 
$BR$ & 0.042 & 0.13 & 0.042 \\ \hline
\end{tabular}
\caption{The $BR$s for $Z'_R$ decay under an approximation assuming $m_{Z'_R}^2 >> m_{f}^2$ where $m_f$ is mass of final state fermion. }
\label{tab:BRs}
\end{table}

{\it $Z'_R$ physics at lepton collider} : Although it would be difficult to produce $Z'_R$ directly at lepton colliders we can explore the effective interaction induced from $Z'_R$ exchange;
\begin{equation}
{\cal L}_{eff} = \frac{1}{1+ \delta_{e \ell}} \frac{g_R^2}{m_{Z'_R}^2} (\bar e \gamma^\mu P_R e)(\bar f \gamma_\mu P_R f), \label{eq:eff}
\end{equation}
where $f$ indicates all the fermions in the model, and only the right-handed chirality appears due to the nature of $U(1)_R$ symmetry.
For example, the analysis of data by LEP experiment in ref.~\cite{Schael:2013ita} provides the constraint $\frac{m_{Z'_R}}{g_R} \gtrsim 3.7 \ {\rm TeV}$.

Furthermore chirality structure of the effective interaction could be tested by measuring the process $e^+ e^- \to f \bar f$ at the International Linear Collider (ILC) using
polarized initial state.
The partially-polarized differential cross section can be defined as~\cite{Nomura:2017abh}
\begin{align}
&\frac{d \sigma (P_{e^-}, P_{e^+})}{d \cos \theta} \nn \\
& = \sum_{\sigma_{e^-}, \sigma_{e^+} = \pm} \frac{1+ \sigma_{e^-} P_{e^-}}{2} \frac{1 +\sigma_{e^+} P_{e^-}}{2} \frac{d \sigma_{\sigma_{e^-} \sigma_{e^+}}}{d \cos \theta},
\end{align}
where $P_{e^-(e^+)}$ is the degree of polarization for the electron(positron) beam and $\sigma_{\sigma_{e^-} \sigma_{e^+}}$ indicates the cross section when the helicity of initial electron(positron) is $\sigma_{e^- (e^+)}$ and the helicity of final states is summed up; more detailed form is found in ref~\cite{Nomura:2017abh}.
The polarized cross sections $\sigma_{L,R}$ is given by following two cases as realistic values at the ILC~\cite{Baer:2013cma}:
\begin{equation}
\frac{d \sigma_{R}}{d \cos \theta} = \frac{d \sigma (0.8,-0.3)}{d \cos \theta}, \quad \frac{d \sigma_{L}}{d \cos \theta} = \frac{d \sigma (-0.8,0.3)}{d \cos \theta}.
\end{equation}
Then we apply $\sigma_R$ to study the sensitivity to $Z'_R$ since it is sensitive to right-handed current interactions~\cite{Nomura:2017tih}. 
To investigate the effect of the new interaction we consider the measurement of a forward-backward asymmetry at the ILC which is given by 
\begin{align}
& A_{FB} = \frac{N_F - N_B}{N_F + N_B}, \nonumber \\
& N_{F(B)} = \epsilon L \int_{0(-c_{\rm max})}^{c_{\rm max}(0)} d \cos \theta \frac{d \sigma}{d \cos \theta},
\end{align}
where a kinematical cut $c_{\rm max} = 0.5(0.95)$ is chosen to maximize the sensitivity for electron(muon)~\cite{Tran:2015nxa}, $L$ is an integrated luminosity and $\epsilon$ is an efficiency depending on the final states which is assumed to be $\epsilon = 1$ for electron and muon final states.
The sensitivity to $Z'_R$ contribution is estimated by 
\begin{equation}
\Delta A_{FB} = |A_{FB}^{SM+Z'_R}- A_{FB}^{SM}|,
\label{eq:delAFB}
\end{equation}
where $A_{FB}^{SM+Z'_R}$ and $A_{FB}^{SM}$ are forward-backward asymmetry for "SM + $Z'_R$" and SM cases respectively.
We compare $\Delta A_{FB}$ with a statistical error of the asymmetry in only SM case 
\begin{equation}
\delta_{A_{FB}}^{SM} = \sqrt{\frac{1-(A_{FB}^{SM})^2}{N_F^{SM}+N_B^{SM}}},
\end{equation}
and we focus on muon final state which is the most sensitive one.
We find that it is difficult to get $\Delta A_{FB} > \delta_{A_{FB}}^{SM}$ for $\sqrt{s} = 250$ or $500$ GeV in the region which satisfy the LHC constraint even if the integrated luminosity is $\mathcal{O}(10)$ ab$^{-1}$.
On the other hand, for $\sqrt{s} = 1$ TeV,  $\Delta A_{FB} \sim 2 \delta_{A_{FB}}^{SM}$ can be obtained with the integrated luminosity of 5 ab$^{-1}$ with $m_{Z'_R}/g_R = 40$ TeV.
Therefore to investigate the chirality structure, we need $\sqrt{s} = 1$ TeV with large integrated luminosity which would be achieved if the ILC is upgraded~\cite{Barklow:2015tja}.

\section{Summary and Conclusions}
We have constructed an inverse seesaw model with $U(1)_R$ symmetry, in which we have formulated the neutrino mass matrix to reproduce current neutrino oscillation data in a general way.
Also we have found a predictive two-zero neutrino mass matrix, by imposing an additional flavor dependent $U(1)_{L_\mu-L_\tau}$ gauge symmetry that also satisfies the gauge anomaly cancellations among $U(1)_R$.
Then we have analyzed collider physics on an extra gauge boson and show a possibility of detection,
Although the result of collider physics is almost the same as the one of our canonical seesaw model~\cite{Nomura:2017tih}.
the neutrino predictions originate from the inverse seesaw model
that could be difficult to reproduce any canonical seesaw models.

\section*{Acknowledgments}
H. O. is sincerely grateful for KIAS and all the members.

\end{document}